\documentclass[12pt,usenames,dvipsnames]{article}

\usepackage{latexsym}
\usepackage{amssymb,amsfonts,amsmath}
\usepackage{graphicx} 
\usepackage{indentfirst}
\usepackage{bbm}
\usepackage{amssymb}
\usepackage{verbatim}
\usepackage{amsmath, amsthm,amssymb}
\usepackage{mathrsfs}
\usepackage{hyperref}
\usepackage{amsfonts}
\usepackage{dsfont}
\usepackage{cite}
\usepackage{xcolor}
\usepackage{enumerate}
\usepackage{cleveref}
\parskip=\medskipamount

\newcommand {\cD}{{\cal D}}
\newcommand {\cE}{{\cal E}}

\newcommand {\cJ}{{\cal J}}

\newcommand {\cN}{{\cal N}}

\def\a{\alpha}

\def\b{\beta}

\def\d{\delta}

\def\g{\gamma}

\def\o{\omega}

\def\q{\theta}

\def\s{\sigma}

\def\z{\zeta}

\def\J{\Psi}

\def\P{\Pi}

\def\U{\Upsilon}

\def\rd{{\rm d}}
\def\ri{{\rm i}}

\newcommand{\ad}{{\dot{\alpha}}}                           %new
\newcommand{\bd}{{\dot{\beta}}}                            %new
\newcommand{\ve}{\varepsilon}                            %new
\newcommand{\cDB}{{\bar\cD}}                            %new

\renewcommand{\aa}{{\a\ad}}

\newcommand{\hf}{\frac12}
\newcommand{\be}{\begin{equation}}
\newcommand{\ee}{\end{equation}}
\newcommand{\bea}{\begin{eqnarray}}
\newcommand{\eea}{\end{eqnarray}}

\newcommand{\bm}[1]{\mbox{\boldmath$#1$}}

\def\double #1{#1{\hbox{\kern-2pt $#1$}}}

\newif\ifdtup

\newcommand{\bsubeq}{\begin{subequations}}
\newcommand{\esubeq}{\end{subequations}}
\numberwithin{equation}{section}

\newcommand{\sSU}{\mathsf{SU}}

\newcommand{\sU}{\mathsf{U}}

\begin{document}

\begin{titlepage}
\begin{flushright}
May, 2023 \\
\end{flushright}
\vspace{5mm}

\begin{center}
{\Large \bf 

The $\cN=2$ superconformal gravitino multiplet

}
\end{center}

\begin{center}

{\bf Daniel Hutchings, Sergei M. Kuzenko and Emmanouil S. N. Raptakis} \\
\vspace{5mm}

\footnotesize{
{\it Department of Physics M013, The University of Western Australia\\
35 Stirling Highway, Perth W.A. 6009, Australia}}  
~\\
\vspace{2mm}
~\\
Email: \texttt{ 
daniel.hutchings@uwa.edu.au, sergei.kuzenko@uwa.edu.au, emmanouil.raptakis@uwa.edu.au}\\
\vspace{2mm}

\end{center}

\begin{abstract}
\baselineskip=14pt

We propose a new gauge prepotential $\Upsilon_i$ describing the four-dimensional ${\cal N}=2$ superconformal gravitino multiplet. The former naturally arises via a superspace reduction of the ${\cal N}=3$ conformal supergravity multiplet. A locally superconformal chiral action for $\Upsilon_i$, which is gauge-invariant in arbitrary conformally flat backgrounds, is derived. This construction readily yields a new superprojector, which maps an isospinor superfield $\Psi_i$ to a multiplet characterised by the properties of a conformal supercurrent associated with $\Upsilon_i$. Our main results are also specialised to ${\cal N}=2$ anti-de Sitter superspace. 

\end{abstract}
\vspace{5mm}

\vfill

\vfill
\end{titlepage}

\newpage
\renewcommand{\thefootnote}{\arabic{footnote}}
\setcounter{footnote}{0}

\tableofcontents{}
\vspace{1cm}
\bigskip\hrule

\allowdisplaybreaks

\section{Introduction}

Two years ago, two of us (SK and ER) proposed models for $\cN=2$ superconformal higher-spin gauge superfields $\U_{\a(m) \ad(n)}$, $m,n \geq 0$ \cite{KR21}.\footnote{The $m=n=0$ case corresponds to the conformal supergravity multiplet \cite{HST} (see also \cite{Siegel-curved,KT} for an explanation of the harmonic-superspace origin of the prepotential $\U$).} The prepotentials $\U_{\a(m) \ad(n)}$ may be deduced by employing an extension of the conformal supercurrent analysis of Howe, Stelle and Townsend \cite{HST}. It was shown that, associated with each gauge superfield, there are two inequivalent\footnote{Actually, if the prepotential is real ($m=n$), the field strengths coincide.} chiral field strengths $\hat{\mathbb{W}}_{\a(m+n+2)}(\U)$ and $\check{\mathbb{W}}_{\a(m+n+2)}(\bar{\U})$, which are generalisations of the linearised super-Weyl tensor. Further, $\hat{\mathbb{W}}^{\a(m+n+2)}(\U)\check{\mathbb{W}}_{\a(m+n+2)}(\bar{\U})$
%their contraction  
constitutes the chiral Lagrangian for $\U_{\a(m)\ad(n)}$ in conformally flat backgrounds \cite{KR21}.

It should be noted that the above linearised super-Weyl tensors 
%above each 
carry at least two spinor indices. Further, the chiral field strength of a massless vector multiplet \cite{GSW} is scalar, $\mathbb{W}(\U)$, where 
$\U_{ij} = \U_{ji} = \overline{\U^{ij}}$ is Mezincescu's prepotential \cite{Mezincescu}. Thus, there is a logical gap in our understanding as the role of the spinor chiral field strengths $\hat{\mathbb{W}}_{\a}(\U)$ and $\check{\mathbb{W}}_{\a}(\bar{\U})$ is unclear. In \cite{KR21} it was postulated that these should correspond to a superconformal gravitino multiplet, though the issue was not explored further. The present work is aimed at bridging this gap.

Our results are primarily formulated within the $\cN=2$ conformal superspace geometry due to Butter \cite{ButterN=2}. The conventions we employ are those of the recent review \cite{KRTM2}.

\section{Superconformal gravitino multiplet}
\label{section2}

In this section we derive the $\cN=2$ superconformal gravitino multiplet by first making a detour to $\cN=3$ conformal supergravity. The motivation for this is due to the latter encoding the former within its multiplet. We begin by uplifting the known construction of the $\cN=3$ conformal supercurrent given in \cite{HST} to general conformally flat backgrounds. In such geometries,
%noting that, in conformally-flat backgounds, 
the $\cN=3$ conformal supercurrent $\mathfrak{J}= (\mathfrak{J}^I{}_J)$ is an Hermitian traceless superfield, 
\begin{align}
	\mathfrak{J}^\dagger = \mathfrak{J} ~, \qquad {\rm tr} \,\mathfrak{J} = 0~,
\end{align}
obeying the conservation equation\footnote{If we restrict our attention to flat superspace, equation \eqref{2.1} reduces to the one proposed in \cite{HST}.}
\begin{align}
	\label{2.1}
	\nabla_{\a}^{(I} \mathfrak{J}^{J)}{}_K = \frac 1 4 \d_K^{(I} \nabla_\a^{|L|} \mathfrak{J}^{J)}{}_{L} \quad \Longleftrightarrow \quad  \bar{\nabla}_{(I}^{\ad} \mathfrak{J}^{J}{}_{K)} = \frac 1 4 \d^J_{(I} \bar{\nabla}^\ad_{|L|} \mathfrak{J}^{L}{}_{K)} ~.
\end{align}
Here $\a$ and $\ad$ are two-component spinor indices,\footnote{
Our two-component spinor notation and conventions follow 
\cite{BK}, and are similar to those adopted in  \cite{WB}.  The only difference is that the spinor Lorentz generators $(\s_{ab})_\a{}^\b$ and 
$({\tilde \s}_{ab})^\ad{}_\bd$  used in \cite{BK} have an extra minus sign as compared with \cite{WB}, 
specifically $\s_{ab} = -\frac{1}{4} (\s_a \tilde{\s}_b - \s_b \tilde{\s}_a)$ and 
 $\tilde{\s}_{ab} = -\frac{1}{4} (\tilde{\s}_a {\s}_b - \tilde{\s}_b {\s}_a)$.  
%Two-component Weyl spinor indices may be raised and lowered in accordance with 
%the rules $\psi^\a = \ve^{\a \b} \psi_\b$, $\psi_\a = \ve_{\a \b} \psi^\b$ and their 
%conjugates. Here $\ve^{\a \b} = - \ve^{\b \a}$, $\ve_{\a \b} = - \ve_{\b \a}$ and their 
%conjugates are the usual $\sSL(2,\mathbb{C})$ invariant tensors. Additionally, we 
%adopt the normalisation $\ve_{2 1} = \ve^{1 2} = 1$.
} 
and capital Latin letters are used to denote $\sSU(3)$ indices, $I, J = \underline{1},\underline{2},\underline{3}$. Our conventions for (conformally flat) $\cN=3$ conformal superspace are those of \cite{KR21}.\footnote{Recently, an off-shell construction of $\cN=3$ conformal supergravity was given in \cite{HMS,MP,HS} within the framework of the superconformal tensor calculus. A superspace analogue of this formulation would prove to be very useful, however it has yet to be constructed. The conformally flat setup developed in \cite{KR21} suffices for our goals in this paper.}
Now, requiring $\mathfrak{J}^{I}{}_J$ to be primary uniquely fixes its dimension and charge as follows:
\begin{align}
	\label{2.2}
	K^M \mathfrak{J}^{I}{}_J = 0 \quad \implies \quad \mathbb{D} \mathfrak{J}^{I}{}_J = 2 \mathfrak{J}^{I}{}_J ~, \quad  \mathbb{Y} \mathfrak{J}^{I}{}_J = 0 ~.
\end{align}
Here $K^M$ denotes the special superconformal generator, while $\mathbb{D}$ and $\mathbb{Y}$ are the dilatation and $\sU(1)_R$ generators, respectively. 

%At this point it is necessary to restrict the background to be conformally-flat. Then, u
Upon performing an $\cN=3 \rightarrow \cN=2$ superspace reduction, $\mathfrak{J}^{I}{}_J$ describes a multiplet of $\cN=2$ conformal supercurrents:
\begin{subequations}
\begin{align}
	\mathcal{J} &:= \mathfrak{J}^{\underline 3}{}_{\underline 3}|_{\q^\a_{\underline{3}} = \bar{\q}_\ad^{\underline {3}} = 0} ~, \label{2.3a} \\
	\mathcal{J}^i &:= \mathfrak{J}^{i}{}_{\underline 3}|_{\q^\a_{\underline{3}} = \bar{\q}_\ad^{\underline {3}} = 0}~, \label{2.3b} \\
	\mathcal{J}^i{}_j &:= \ri \mathfrak{J}^{i}{}_{j}|_{\q^\a_{\underline{3}} = \bar{\q}_\ad^{\underline {3}} = 0} + \frac{\ri}{2} \d^i_j \mathcal{J} ~, \label{2.3c}
\end{align}
\end{subequations}
where $i,j = \underline{1}, \underline{2}$ are $\sSU(2)$ indices.\footnote{These isospinor indices are raised and lowered by using the antisymmetric $\sSU(2)$-invariant tensor $\ve_{ij}= -\ve_{ji}$ and its inverse $\ve^{ij}=-\ve^{ji}$, normalised as $\ve_{\underline{2} \underline{1}} = \ve^{\underline{1} \underline{2}} = 1$, via the rule $\chi^i = \ve^{i j} \chi_j$, $\chi_i = \ve_{i j} \chi^j$. } Here $\mathcal{J}$ corresponds to the ordinary conformal supercurrent \cite{HST,Sohnius,KT}, $\mathcal{J}^{ij} = \mathcal{J}^{ji}$ is the linear multiplet \cite{BS,SSW} and $\mathcal{J}^i$ is a new supercurrent which necessarily couples to the superconformal gravitino multiplet. Making use of \eqref{2.1}, we see that $\mathcal{J}^i$ obeys the conservation equations:
\begin{align}
	\label{2.4}
	{\nabla}_\a^{(i} \mathcal{J}^{j)} = 0 ~, \qquad \bar{\nabla}^{(ij} \mathcal{J}^{k)} = 0 ~,
\end{align}
and is characterised by the superconformal properties:
\begin{align}
	K^M \mathcal{J}^i = 0 \quad \implies \quad \mathbb{D} \mathcal{J}^i = \mathbb{Y} \mathcal{J}^i = 2 \mathcal{J}^i ~.
\end{align}
Here we have defined:
\begin{align}
\nabla^{ij} = \nabla^{\a (i} \nabla_\a^{j)} ~, \qquad \bar{\nabla}_{ij} = \bar{\nabla}_{\ad (i} \bar{\nabla}^\ad_{j)} ~.
\end{align}
It should be emphasised that, owing to the complex nature of $\cJ^i$, $\bar{\nabla}^{\ad (i} \cJ^{j)} \neq 0$ in general.
A simple realisation of the present multiplet is $\mathcal{J}^i = \bar{\mathbb{W}} q^i$, where $\mathbb{W}$ is the chiral field strength of a $\cN=2$ vector multiplet and $q^i$ describes an on-shell hypermultiplet.

We now employ the method of supercurrent multiplets \cite{HST,OS,BdRdW} to determine the gauge prepotential associated with $\mathcal{J}^i$, which we denote $\U_i$.\footnote{The complex conjugate of $\U_i$ is defined as $\bar{\U}^i := \overline{\U_i}$. Hence, it follows that $\overline{\U^i} = - \bar{\U}_i$ .} Specifically, the latter may be determined by requiring that the Noether coupling
\begin{align}
	\label{2.5}
	\mathcal{S}_{\text{N.C.}} = \int \rd^4x \rd^4 \q \rd^4 \bar\q \, E\, \U_i \, \mathcal{J}^i + \text{c.c.} ~,
\end{align}
is locally superconformal and gauge-invariant. Here $E$ is the integration measure for the full superspace. It is clear to see that $\U_i$ is defined modulo the gauge freedom\footnote{It should be emphasised that the gauge transformation law \eqref{2.6} is valid in generic supergravity backgrounds.}
\begin{align}
	\label{2.6}
	\d_{\z, \o} \U_i = \nabla^{\a j} \z_{\a ij} + \bar{\nabla}^{jk} \o_{ijk}~,
\end{align}
and its superconformal transformation law is characterised by the properties:
\begin{align}
	\label{2.7}
	K^M \U_i = 0 ~, \qquad \mathbb{D} \U_i = \mathbb{Y} \U_i = -2 \U_i~.
\end{align}
It should be emphasised that the superconformal properties \eqref{2.7} imply that the gauge transformations \eqref{2.6} are superconformal. Further, the gauge superfield $\U_i$ describes the $\cN=2$ superconformal gravitino multiplet.

Having determined the new gauge prepotential $\U_i$ above, it remains to determine its kinetic action. For simplicity, we perform this analysis in conformally flat superspace backgrounds. In particular, one may show that the following chiral descendants
\begin{align}
	\label{2.8}
	\hat{\mathbb{W}}_\a(\U) := \bar{\nabla}^{4} \nabla^{ij} \nabla_{\a i} \U_j ~, \qquad \check{\mathbb{W}}_\a(\bar{\U}) := \bar{\nabla}^{4} \nabla_\a^i \bar{\U}_i ~, \qquad \bar{\nabla}^4 = \frac{1}{48} \bar{\nabla}^{ij} \bar{\nabla}_{ij}~,
\end{align}
are gauge-invariant and are characterised by the superconformal properties:
\begin{subequations}
	\label{2.9}
\begin{align}
	K^M \hat{\mathbb{W}}_\a(\U)  &= 0 ~, \qquad \mathbb{D} \hat{\mathbb{W}}_\a(\U)  = \frac{3}{2} \hat{\mathbb{W}}_\a(\U)  ~, \qquad \mathbb{Y} \hat{\mathbb{W}}_\a(\U)  = - 3 \hat{\mathbb{W}}_\a(\U) ~, \\
	K^M \check{\mathbb{W}}_\a (\bar{\U}) &= 0 ~, \qquad \mathbb{D} \check{\mathbb{W}}_\a (\bar{\U}) = \frac{1}{2} \check{\mathbb{W}}_\a (\bar{\U}) ~, \qquad \mathbb{Y} \check{\mathbb{W}}_\a (\bar{\U}) = - \check{\mathbb{W}}_\a (\bar{\U})~.
\end{align}
\end{subequations}

It then follows that
\begin{align}
	\label{2.10}
	\mathcal{S}_{\text{SCGM}}[\U,\bar{\U}] = \frac{1}{2} \int \rd^4x \rd^4 \q \, \cE\, \hat{\mathbb{W}}^\a (\U)   \check{\mathbb{W}}_\a (\bar{\U}) + \text{c.c.} ~,
\end{align}
is the locally superconformal and gauge-invariant action for the $\cN=2$ superconformal gravitino multiplet in conformally flat backgrounds. Here $\cE$ is the chiral integration measure. The overall coefficient in \eqref{2.10} has been chosen due to the identity
\begin{align}
	\label{2.11}
	\ri \int \rd^4x \rd^4 \q \, \cE\, \hat{\mathbb{W}}^\a (\U)   \check{\mathbb{W}}_\a  (\bar{\U}) + \text{c.c.} = 0~,
\end{align}
which holds up to a total derivative.

Upon integration by parts, the action \eqref{2.10} may be written in the equivalent forms:
\bsubeq \label{2.14}
\begin{align}
\mathcal{S}_{\text{SCGM}}[\U,\bar{\U}] &= \hf \int \rd^4x \rd^4 \q \rd^4 \bar{\q} \, E \, \bar{\U}^i \hat{\mathbb{B}}_i (\U)  + \text{c.c.} ~ \label{2.15a} \\
&= \hf \int \rd^4x \rd^4 \q \rd^4 \bar{\q} \, E \, \bar{\U}^i \check{\mathbb{B}}_i (\U)  + \text{c.c.} \label{2.15b}
\end{align}
\esubeq
Here, $\hat{\mathbb{B}}_i (\U) $ and  $\check{\mathbb{B}}_i (\U)$ may be understood as analogues of the linearised super-Bach tensor for the superconformal gravitino multiplet
\bsubeq \label{BachTensors}
\begin{align} 
\hat{\mathbb{B}}_i (\U) &:= \nabla^{\a }_i \hat{\mathbb{W}}_\a (\U)~, \label{BachTensor1} \\
\check{\mathbb{B}}_i (\U) &:=  \bar{\nabla}^{\ad j} \bar{\nabla}_{ij} \bar{\check{\mathbb{W}}}_\ad (\U) ~. \label{BachTensor2}
\end{align}
\esubeq
They are gauge-invariant and satisfy the constraints
\bsubeq
\begin{align}
\label{2.13}
\nabla_{(ij} \hat{\mathbb{B}}_{k)} (\U)&= 0 ~, \qquad \bar{\nabla}_{\ad (i} \hat{\mathbb{B}}_{j)} (\U) = 0~, \\
\nabla_{(ij} \check{\mathbb{B}}_{k)} (\U) &= 0 ~, \qquad \bar{\nabla}_{\ad (i} \check{\mathbb{B}}_{j)} (\U)  = 0~.
\end{align}
\esubeq
By virtue of these properties, the functionals \eqref{2.14} are manifestly gauge-invariant. Additionally, the relations \eqref{2.15a} and \eqref{2.15b}, which 
%by employing 
follow from
the identity \eqref{2.11}, imply that the super-Bach tensors $\hat{\mathbb{B}}_i (\U) $ and  $\check{\mathbb{B}}_i (\U)$ 
%can be shown to 
coincide, 
\begin{align}
	\label{2.12}
	\mathbb{B}_i(\U) := \hat{\mathbb{B}}_i(\U) = \check{\mathbb{B}}_i (\U)  ~.
\end{align}

%where $\mathbb{B}^i$ may be understood as an analogue of the super-Bach tensor for the superconformal gravitino multiplet. Routine calculations lead to the differential constraints

%which, owing to the constraints \eqref{2.13}, is manifestly gauge-invariant. 

%It is clear that the equation of motion obtained by varying \eqref{2.14}  with respect to $\bar{\U}_i$ is 
%\begin{align}
%\hat{\mathbb{B}}_i (\U)  = \check{\mathbb{B}}_{i} (\U) = 0~.
%\end{align}

\section{Superprojectors}
\label{section3}
Superprojectors \cite{SalamStrathdee, Sokatchev1975, Sokatchev1981, RittenbergSokatchev1981, SiegelGates1981, GatesGrisaruRocekSiegel1983, Buchbinder2019, Buchbinder2021, Hutchings2021} are non-local idempotent differential operators which extract out the component of a superfield which realises an irreducible representation of the corresponding supersymmetry algebra. Such operators have found extensive applications within the literature, including in: i) the computation of field equations which describe massive superfields \cite{Ogievetsky1976qc,Ogievetsky1976qb}; ii) the formulation of gauge-invariant actions which describe massless superfields \cite{Gates1979gv, Gates2003cz}; and iii) the construction of linearised actions for superconformal higher-spin multiplets \cite{Buchbinder2019, Buchbinder2021, Hutchings2021}. In this section, we make use of the action \eqref{2.14} describing the $\cN=2$ superconformal gravitino multiplet to construct the projection operators\footnote{The relationship between irreducible representations of the four-dimensional $\cN=2$ superconformal algebra $\mathfrak{su}(2,2|2)$ and the superprojectors \eqref{3.4} will be studied elsewhere.} which map an unconstrained isospinor superfield $\Psi_i$ characterised by the superconformal properties\footnote{We emphasise that a general unconstrained isospinor $\bm{\Psi}_i$ may always be mapped to one obeying \eqref{3.1} by utilising some conformal compensator(s).}
\begin{align}
	\label{3.1}
	K^M \Psi_i = 0 ~, \qquad \mathbb{D} \Psi_i = \mathbb{Y} \Psi_i = -2 \Psi_i~,
\end{align}
into a multiplet which has the properties \eqref{2.4} of a conformal supercurrent.

One can immediately introduce the differential operators $\hat{\mathbb{P}}_i{}^j$ and $\check{\mathbb{P}}_i{}^j$, which are defined by their action on $\Psi_i$
\vspace{-\baselineskip}
\bsubeq
\begin{align}
\hat{\mathbb{P}}_i{}^j \Psi_j &:= \phantom{-} \nabla^\a_i \bar{\nabla}^4 \nabla^{jk}\nabla_{\a j} \Psi_k~, \label{N2BachOperator1} \\
\check{\mathbb{P}}_i{}^j \Psi_j &:= - \bar{\nabla}^{\ad k} \bar{\nabla}_{ik} \nabla^4 \bar{\nabla}_\ad^j \Psi_j ~. \label{N2BachOperator2} 
\end{align}
\esubeq
Note that $\hat{\mathbb{P}}_i{}^j \Psi_j$ and $\check{\mathbb{P}}_i{}^j \Psi_j $ take the same structural form as the super-Bach tensors \eqref{BachTensor1} and \eqref{BachTensor2}, respectively.
However, the operators \eqref{N2BachOperator1} and \eqref{N2BachOperator2} are not idempotent
\bsubeq  \label{N2BachOperatorSquared}
\begin{align}
\hat{\mathbb{P}}_i{}^j \hat{\mathbb{P}}_j{}^k \Psi_k = -24 \Box^2 \hat{\mathbb{P}}_i{}^k \Psi_k~, \\
\check{\mathbb{P}}_i{}^j \check{\mathbb{P}}_j{}^k \Psi_k = -24 \Box^2 \check{\mathbb{P}}_i{}^k \Psi_k~,
\end{align}
\esubeq
where we have defined $\Box := -\hf \nabla^{\a\ad}\nabla_{\a\ad}$. 

Studying eq. \eqref{N2BachOperatorSquared}, it is easy to see that one can introduce the operators $\hat{\P}_i{}^j$ and $\check{\P}_i{}^j $, which act on $\Psi_i$ in the following manner
\bsubeq
\label{3.4}
\begin{align}
\hat{\P}_i(\Psi) &:= \hat{\P}_i{}^j \Psi_j = -\frac{1}{24 \Box^2} \hat{\mathbb{P}}_i{}^j \J_j = -\frac{1}{24 \Box^2}\nabla^\a_i \bar{\nabla}^4 \nabla^{jk}\nabla_{\a j} \Psi_k~, \label{N2Projector1}\\
\check{\P}_i(\Psi)&:= \check{\P}_i{}^j \Psi_j :=  -\frac{1}{24 \Box^2} \check{\mathbb{P}}_i{}^j \J_j = \phantom{-} \frac{1}{24 \Box^2} \bar{\nabla}^{\ad k} \bar{\nabla}_{ik} \nabla^4 \bar{\nabla}_\ad^j \Psi_j  ~. \label{N2Projector2}
\end{align}
\esubeq  
They are, by construction, idempotent
\bsubeq \label{ProjIdemp}
\begin{align}
\hat{\P}_i{}^k \hat{\P}_k{}^j \Psi_j &= \hat{\P}_i{}^j \Psi_j ~, \\
\check{\P}_i{}^k \check{\P}_k{}^j \Psi_j &= \check{\P}_i{}^j \Psi_j ~.
\end{align}
\esubeq
It follows immediately from eq. \eqref{2.12} that the projected superfields \eqref{N2Projector1} and \eqref{N2Projector2} coincide
\be \label{ProjEquiv}
\hat{\P}_i(\Psi) = \check{\P}_i(\Psi)~.
\ee
In accordance with this equivalence, we will only consider the operator $\P_i{}^j := \hat{\P}_i{}^j$ in the subsequent discussion.

By construction, the operator ${\P}_i{}^j$  projects any superfield $\Psi_j$ to one satisfying the same differential constraints as the super-Bach tensor
\eqref{BachTensors}
\be \label{ProjProp1}
\nabla_{(ij} {\P}_{k)}(\Psi) = 0 ~, \qquad \bar{\nabla}_{\ad (i} {\P}_{j)}(\Psi) = 0~.
\ee
Furthermore, ${\P}_i{}^j$ acts as the identity on the space of conformal supercurrents \eqref{2.4}
\be \label{ProjProp3}
\bar{\nabla}_{\ad ( i }\Psi_{j)} = 0~, \quad \nabla_{(ij}\Psi_{k)} = 0 \qquad \Longrightarrow \qquad \P_i(\Psi) = \Psi_j ~.
\ee
In order to show this explicitly, it proves useful to make use of the following identity \cite{SiegelGates1981}\footnote{Each of the terms appearing on the right hand side of \eqref{FMN2ResIde} are themselves orthogonal projectors when acting on the space of scalar fields.}
\be \label{FMN2ResIde}
\Box^2 = \bar{\nabla}^4 \nabla^4 + \nabla^4 \bar{\nabla}^4 + \frac{1}{16} \nabla^{ij} \bar{\nabla}^4  \nabla_{ij} - \frac{1}{16} \nabla^{\b \g} \bar{\nabla}^4 \nabla_{\b \g} - \frac{1}{12} \nabla^{\b i}  \bar{\nabla}^4 \nabla_{i j } \nabla_\b^j - \frac{1}{12} \nabla^{\b i} \nabla_{i j }   \bar{\nabla}^4 \nabla_\b^j ~,
\ee
where we have introduced the notation $\nabla_{\b \g} = \nabla^j_{(\b} \nabla_{\g ) j}$. Thus, the operator $\P_i{}^j $ is a superprojector  which maps any superfield $\Psi_i$ to a conformal supercurrent \eqref{2.4}, as a consequence of the properties \eqref{ProjIdemp}, \eqref{ProjProp1}  and \eqref{ProjProp3}.  In the flat superspace limit, $\nabla_A \rightarrow D_A$, the superprojector \eqref{3.4} may be obtained 
%If one considers the flat superspace limit of our superprojector, $\nabla_A \rightarrow D_A$, it coincides with the one which may be obtained
% Although they were not derived explicitly, the superprojectors can be obtained 
via the prescription described by Rittenberg and Sokatchev  \cite{RittenbergSokatchev1981} or Siegel and Gates in \cite{SiegelGates1981}. The construction of the operator \eqref{3.4} initiates the study of superprojectors which act on superfields carrying isospinor indices in $\cN=2$ conformal superspace.

%As an application of the superprojector \eqref{3.4}, we can recast the superconformal action \eqref{2.10} in the following form
%\be
%\mathcal{S}_{\text{SCGM}}[\U,\bar{\U}] = -12  \int \rd^4x \rd^4 \q \rd^4 \bar{\q} \, E \, \bar{\U}^i \Box^2 {\P}_i (\U)  + \text{c.c.} 
%\ee

%%%%%%%%%%%%%%%%%%%%%%%%%%%%%%%%%

\section{Degauging to $\text{AdS}^{4|8}$}

As we have seen above, the formalism of conformal superspace significantly simplifies calculations in conformally flat backgrounds. However, when considering applications, it is often more useful to work with Lorentz covariant derivatives as opposed to their conformal counterparts. There is a systematic procedure, known as degauging \cite{ButterN=2}, to translate results from conformal superspace to the $\sU(2)$ superspace of \cite{Howe}. While the general scheme of degauging is well-known, it is often highly non-trivial to perform on generic curved backgrounds, but is significantly simpler when several torsions are set to zero.

To this end, we now degauge our main results to $\text{AdS}^{4|8}$.\footnote{It was proven in \cite{BILS} that four-dimensional $\cN$-extended AdS superspace, $\text{AdS}^{4|4\cN}$, is conformally flat.} We recall that its geometry is described by the covariant derivatives $\cD_A = (\cD_a, \cD_\a^i, \bar \cD^\ad_i)$, obeying the algebra (see, e.g. \cite{KT-M-ads}):
\be
\{ \cD_\a^i , \cD_\b^j \} = 4 S^{ij} M_{\a \b} + 2 \ve_{\a \b} \ve^{i j} S^{kl} J_{kl} ~, \quad \{ \cD_\a^i , \bar \cD^\bd_j \} = - 2 \ri \d_j^i \cD_{\a}{}^{\bd} ~,
\ee
where $M_{\a \b}$ and $J_{ij}$ are the Lorentz and $\sSU(2)_R$ generators, respectively. Additionally, the torsion ${S}^{ij} $ is subject to the constraints
\begin{align}
	\label{3.2}
	\cD_A S^{jk} = 0 ~, \qquad {S}^{ij} = { S}^{ji}~, \qquad [S, S^\dagger ]=0~,
\end{align}
where we have defined $S= (S^i{}_j)$. The final constraint in \eqref{3.2} implies that $S^{ij}$ can be chosen to be real, $\overline{S^{ij}} = S_{ij}$, however we will not impose this condition.

We first derived the conformal supercurrent $\mathcal{J}^i$, which was subject to the differential constraints \eqref{2.4}. Upon degauging, the latter take the form:
\begin{align}
	\cD_\a^{(i} \cJ^{j)} = 0 ~, \qquad \big( \bar{\cD}^{(ij} + 4 \bar{S}^{(ij} \big) \cJ^{k)} = 0~,
\end{align}
where $\bar{\cD}^{ij} = \bar{\cD}_\ad^{(i} \bar{\cD}^{\ad j)}$ and $\cD^{ij} = \cD^{\a (i} \cD_{\a}^{j)}$. It is dual, via the Noether coupling \eqref{2.5}, to the gauge prepotential $\U_i$, which is defined modulo the gauge transformations
\begin{align}
	\d_{\z,\o} \U_i = \cD^{\a j} \z_{\a ij} + \big( \bar{\cD}^{jk} + 4 \bar{S}^{jk} \big) \o_{ijk}~,
\end{align}
where $\z_{\a ij}$ and $\o_{ijk}$ are complex unconstrained. The gauge-invariant field strengths associated with $\U_i$ (and its conjugate $\bar{\U}^i$) are:
\begin{align}
	\label{4.5}
	\hat{\mathbb{W}}_{\a}(\U) &=  \bar{\cD}^4 (\cD^{ij} + 6 S^{ij}) \cD_{\a i} \U_j ~, \quad
	\hat{\mathbb{W}}_{\a}(\bar{\U}) = \bar{\cD}^4 \cD_\a^i \bar{\U}_i ~, \quad \bar{\cD}^4 = \frac{1}{48}\big(\bar{\cD}^{ij} + 4 \bar{S}^{ij} \big) \bar{\cD}_{ij}~.
\end{align}
They are utilised to construct the kinetic action for $\U_i$, see eq. \eqref{2.10}. 

Further, from the AdS-specific field strengths \eqref{4.5}, we derive the linearised super-Bach tensor \eqref{BachTensor1} in this geometry
\begin{align}
	\label{AdSBach}
	\mathbb{B}_i (\U) = \cD^\a_i \hat{\mathbb{W}}_{\a}(\U) = \cD^\a_i \bar{\cD}^4 \big(\cD^{jk} + 6 S^{jk}\big) \cD_{\a j} \U_k~,
\end{align}
which is manifestly gauge-invariant and satisfies the constraints:
\begin{align}
	\label{AdSBachConstraints}
	\big(\cD_{(ij} + 4 S_{(ij}\big) \mathbb{B}_{k)} = 0 ~, \qquad \cDB_{\ad(i} \mathbb{B}_{j)} = 0~.
\end{align}

In section \ref{section3}, by utilising the super-Bach tensor as a prototype, we proposed a new superprojector $\Pi_i{}^j$, which maps an unconstrained isospinor $\Psi_i$ to the space of superfields obeying the constraints \eqref{ProjProp1}.\footnote{The explicit form of the superprojectors which act on the space of unconstrained superfields $\J_{\a(m)\ad(n)}$ in AdS$^{4|8}$, for $m, n \geq 1$, were recently derived in \cite{Hutchings2023}. } In accordance with the analysis above, this operator takes the following form in $\text{AdS}^{4|8}$
\begin{align}
	\label{4.8}
	\Pi_i(\Psi) = -\frac{1}{24 \mathbb{Q}^2} \cD^\a_i \cDB^{4} \big(\cD^{jk} + 6 S^{jk}\big) \cD_{\a j} \Psi_k~.
\end{align}
Here $\mathbb{Q}$ is the quadratic Casimir operator of the $\cN=2$ AdS superalgebra $\mathfrak{osp}(2|4)$ \cite{Hutchings2023}
\be
\label{AdSCasimir}
\mathbb{Q} : = - \hf \cD^{\aa} \cD_{\aa} - \frac{1}{4} \big(\bar{S}^{ij} \cD_{ij} + S^{ij }\cDB_{ij} \big ) -\hf S^{ij} \bar{S}_{ij} \big (M^{\a \b} M_{\a \b} +\bar{M}^{\ad \bd} \bar{M}_{\ad \bd} \big ) + \hf S^{ij} \bar{S}^{kl}J_{ij}J_{kl}~,
\ee
which is characterised by the property $[\mathbb{Q},\cD_A]=0$. It may be shown that \eqref{4.8} projects onto the space of superfields satisfying the constraints
\begin{align}
	\big(\cD_{(ij} + 4 S_{(ij}\big) \Pi_{k)}(\Psi) = 0 ~, \qquad \cDB_{\ad(i} \Pi_{j)}(\Psi) = 0~.
\end{align}

\section{Conclusion}

In this paper, we proposed a new conformal supercurrent multiplet $\cJ^i$, characterised by the constraints \eqref{2.4}, and its associated gauge prepotential $\U_i$, which describes the $\cN=2$ superconformal gravitino multiplet. Building on this, we construct the model for $\U_i$ on general conformally-flat backgrounds, completing the research program initiated in \cite{KR21}. From this action, we then construct a new superprojector which maps any isospinor superfield $\J_i$ to one satisfying \eqref{ProjProp1}. 

%%%%%%%%%%%%%%%%%%%%%%%%%%%%%%%%%

\noindent
{\bf Acknowledgements:}
We are grateful to the referee of this paper for useful suggestions. This work was supported in part by the Australian Research Council, project No. DP200101944 and DP230101629. DH and ER thank the Mathematical Research Institute MATRIX in Australia for hospitality and support during the early stages of this project.

\begin{footnotesize}

\end{footnotesize}

\end{document}